# Parametric Bayesian Rejuvenation in Ambient Assisted Living through Software-based Thematic 5G Management

Rossi Kamal, Choong Seon Hong
Department of Computer Engineering
Kyung Hee University, South Korea
Email: rossikamal@rossikamal.info,cshong@khu.ac.kr

## ABSTRACT

Ameliorating elderly engagement is vital in rejuvenating independent living. However, recommended practices lack realization of personal traits despite socio -economic promise. The recent proliferation of IoT with the advent of smart-objects/things and personalized services pave the way for context-aware service management. Eventually, the major goal of this paper is to develop a context-aware model in predicting engagement of elderly care. Hence, key requirements are identified for elderly engagement, namely (a) discovery of contexts, which are relevant (b) scaling up (over time) of engagement. However, paramount challenges are imposed on this stipulation, such as, un-observability, independence and composite relationship of contexts. Therefore, a Topic-model based model is proposed to address scalability of contexts and its conjugal relationship with engagement. Eventually, systematic framework is demonstrated, which pinpoints key goals of context-aware services by participants' opinions, usage and feed-back.

## INTRODUCTION

United Nations (UN)s Population Division anticipates that senior citizens (aged 65 or above) might exceed young people (aged 20 to 64) for the first time in history around 2050[1]. This ratio of senior to young would exceed 30% in 2020 and reach 40% by 2030. Moreover, worldwide the senior population has tripled in last 50 years and is expected to be tripled in the next 50-years. This is originated from the significant higher annual growth (1.9%) of the senior citizen (1.9%) than that (1.02%) of mass-people. In the United States (US), the senior citizens, who were born in World War II, represent largest community (i.e 76 million). On the other hand, European Commission (EC) predicts a 44% rise of seniors between 1995 and 2025 in only UK.

In this context, possible social-economic confluence of aging generations on developed countries are striving policy makers and utilitarian ages towards scientific endeavors[2][3][4][4][5][6][7][8] for improving overall urban life. Elderly compensates a giant amount on health insurance from the wealth earned through his entire career. However, they long for reduction in dependency on others, even compromising some comfort for wearable environment. Evolved from aging factors, an advanced technology is demanded, which is, ambient with pervasive computation, autonomous with self-managed modules, and most importantly, accessible, opera- tional and permissible for seniors. Initially this demand earned adequate attention around 80s with simple activity-monitoring Fig. 1. South Korean version of Ambient Assisted Living[6] tools, which stumbles onward through the recent implication of Smart wearable and in-vivo devices[4][4][5][6][7].

Eventually, scientific methodology gradually emerges from closed-door laboratory to miniature-scaled[9] connected-home deployments. Several initiatives turn popular around the world with similar slogans Aging Well and Prolonging Indepen- dent Living(Fig. 1)[2][3][4][4][5][6][7][8]. Ambient Assisted Living (AAL)[3] is European technology and innovation programme funded by European Commission and also syn- chronized with FP7 research area. The slogan is given a dimension with significance of recently launched program entitled Horizon 2020[2]. Acquiring recommended practices, such as, electronic health (e-health), mobile health (m-health), and smart health (s-health), AAL is renovated with ICT technology to sense information from objects, or worn on elderly or surrounding environments in autonomous regulation of well-being for an independent life. The overall goal is to develop intelligent products and services with physical equipments (i.e sensing, processing, transmission devices), which are connected using various networked and software elements. All funding programs promote independent elderly living towards a overall healthier society.

Participants' engagement on ambient sensors is obligatory because of responsiveness to multi-modal contexts[10][11],especially behavior, activity or even demand. There exist things or objects[9], which resemble surrounding contexts of citizens and represent unusual or sudden change of aging or psychological factors. Thus, adequate reasoning and inference of participation information govern caregivers' optimized decision making. Therefore. customized and user-friendly inter- faces and services are compulsory to overcome participants' impairments, especially in mobility, hearing, vision, etc. Such participatory behavior seldom become intricate in small scale deployment, such as simple care. However, engagement is pivotal in production-scale deployments, which necessitates detection of vital ambient sign and thereby inference of common and unusual pattern from knowledge base as primary concerns. Hence, engagement proliferates among multidisciplinary research communities, such as healthcare, artificial intelligence, sensing, communication, cognition, privacy, ethics, psychology, etc.

Recently, 5G vision[12] is emerging with the vision of faster data[13], stringent latency, optimal balance of cost and performance[14][15][16]. With the advent internet-of-Things (IoT)/M2M(Machine-to-Machine Communication)[4][9]-enabled Smart-phones, wearable de- vices and autonomous vehicles, 5G is expected to pave the way for Smarter urban living, where personalization[17] is to be inaugurated by reasoning underlying networked environment to expedite superb stimuli of experience for consumers. The increasing craving for connectivity and customize services promises to renovate service ecosystem and thereby expands plethora of business-cases of internet community. Having strong potential impact on both formal and informal spaces, such as quality of experience and urban-provisioning, technology is being considered as key player in bringing Smartness in urban community. Eventually, intelligence, sen- sitivity and responsiveness embedded in things or objects, or ubiquitous surrounding, are demanding policy-based and autonomous[18],[19],[20] redefinition of urban space, ecosys- tem, and consequent approaches[21],[22],[23].

In this context, enabling participation information[9] as a utility promises to improve independent living[3]. It includes exploration of value in massive data, which is collected from heterogeneous sources, such as public (i.e. hospital), private (e.g. Smart-home) agencies or social presence (i.e. Facebook or Twitter) or regular-usage (i.e. YouTube). However, not only geography, but also other essences of data, such as mobility, consumer-behavior, usage-hour (peak/off-peak), are regarded important. Uncertainties and privacy concerns over data from heterogeneous sources necessitate high control over overall monitoring information. In this context, analytic is indispensable for unwrapping underlying structure for optimal clinical decision making. Thereafter, ambient assisted living is coined by Smart-living, resembled by on-demand provisioning of engagement for both regulatory and emergency services in completely automotive manner.

The objective of the work is to develop instinctive instrumentation, wherefore engagement is stimulated with accurate quantification of contexts. Hence, key prerequisites are speck- led, which automation should clinch. Primarily, a discovery model is essential, which accurately annotates latent theme from plethora of elderly information. However, the latter stage demands scalability, which illuminates transformation over time and attachment on connected themes. This provides both ease and convenience for remote caregivers in optimized decision making, so that elderly enjoy independent daily- living.

To encounter above confrontations, our Topic-model based parametric Bayesian[24] solution Yuva predicts an autonomous model for uncovering themes and scaling up in indepen- dent elderly-engagement. Medium-scale measurements (i.e. subjective, objective-based and acceptability assessments) are conducted to quantify statistical methodology in inferring polychromatic experiences of participants. However, an im- practical implication of statistical information, may lead to conclusion, that is not apotheosis. Therefore, our developed platform[9] attempts to grapple with some completely baffling ingredients.

A synopsis of paper is as follows, (a) a methodical spot line of critical complications of rejuvenation problem in AAL, (b) topic-model based confrontation of discovery and engagement of senior citizens, (c)medium-scale assessment on implication of developed scheme with practical testimony of prototype implementation. (e)concluding remarks.

**PROBLEM FORMULATION**

**A. Ambient Assisted Living Perspective**
ICT activities of Horizon 2020, Ambient Assisted Living and FP7 promote innovative solutions to acquire market share, especially with SMEs and business objectives. The market oriented strategies (e.g. elderly care, smart-home management, awareness program, online promotions) envision collabora- tions of industry and academia at both international and national levels among EU member states. Elderly care monitoring involve daily activities, especially at occurrence of any unusual health-hazards. All endeavors, such as activity monitoring or life-logging necessitate in- teraction of elderly, so that caregivers make optimized res- cue actions on security alerts. Moreover, smart-home based monitoring aim overall quality of life with self-management technologies. These projects demand communication through embedded devices, ambient social networks and elder-friendly customized interfaces. However, awareness programs guide elderly in indoor and outdoor vicinities, such as hospitals, shopping complex and public transports. These actions appear as live-saver for elderly with cognitive impairments. However, adequate activity recognition methodologies are prerequisites for these special services. Last but not the least, online promo- tions stimulate elderly interaction with community broadcast platforms, such as TV, social network, augmented gaming and virtual mentoring. These ventures need elderly participation for loneliness-free and enjoyable independence.

**B. 5G Management Perspective**
5G is proliferating with the promise of faster data, stringent latency, privacy-aware availability, optimal balance between cost and performance. Iniatives are undergone with '5G Infras- tructure Public Private Partnership (5G PPP) in Europe, 2020 and beyond Ad Hoc group in Japan and 5G forum formed in South Korea, OST (Ministry of Science & Technology) 5G project in China. Operators are commonly seen migrating to SDN,NFV and cloud-enabled solutions due to the changed circumstances for high bandwidth and personalized service requirements. New and unseen challenges are frequently im- posed on communication ecosystem, especially with high- bandwidth scavenging multimedia applications. Specially, new methodologies are felt to be implicated in provisioning dy- namic resource and establishing parallelism among end-user requirements. In this context, there is a common trend among operators to connect every 'things' or 'objects' around humans as objects with the form of Internet-of-Things /Machine-to- Machine Communication. To inaugurate digital presence of smart objects around people, a proper realization is necessary about humans and their connectivity requirement, like how, when, where they are generally connected. This renovation with context-aware digital 5G revolution is expected to change the urban lifestyle. Eventually, design of innovative behavior and context-aware services are recommended in 5G manage- ment paradigms. Moreover, the increasing demand of person- alized service is demanding renovation in different steps of value-chain, in content production and delivery. In a nutshell, improving overall consumer perception is emerging among operators and service providers. They are especially concerned about quality of experience at consumer-end, after setting up network infrastructure. However, the convectional procedure is subjective and intricate due to the involvement of domain- specific knowledge and human-intervention, which cost both time and money. On the other hand, consumers will only pay high price for network, if they are satisfied for networked service, different context of service and conveniences of using that service. Hence, system design should integrate adaptation to activity at service-consumption or multivariate contexts, provisioning of performance degradation of experience level and quality and validation of consumers expectation. Hence, migrating tenants from psychology, network-economics, sta- tistical learning tools are becoming popular in management practices to predict user-centric 5G management, evaluation and control.

In a nutshell, the overall objective is to instrument system- atic procedure, such that participation is stimulated with ac- curate quantification of elderly contexts, for example, emotion (i.e. especially for cognitive impaired), location (e.g. hospital, smart home, etc.), social presence (e.g. community TV, social network, etc.).

However, uncovering common contexts from plethora of participation information (Revelation problem) is a mammoth task. Naive approach is to label content with caregivers' own choices and thereby seek for elderly feedback/usage-history on that. However, it involves manual configuration and human involvement, which are time and money-sensitive.

Moreover, representation of huge content (Appurtenance Problem) based on annotated contexts, which are hidden inside, is also cumbersome. The methodology necessitates observation, whether content matched to that contexts, which are changed over time and connected to each other.

**PROPOSED SCHEME**
Speculation on independent-living manifests an oblivious anatomy, which context-awareness should clinch. Elemen- tary stage should entail disclosure of annotated theme from plethora of participation information. However, succeeding stage should expedite transformation over time and composite relations among contexts. Thus, an autonomous model is perceived, which uncovers contexts and scales up with time. Eventually, topic-model based algorithm Yuva is proposed, which tailors contexts, that pervade unstructured but massive collection of elderly information. Hence, Yuva discovers and annotates content with latent thematic information, which archanged over time and synchronized to each other. Yuva is decomposed into three phases, namely espouse, divulge and tryst. Content is initially distributed randomly with fixed patterns (espouse). Usually corpora contains multiple themes (i.e. resembling contexts). However, each corpora exhibits themes in different proportions. The latter phases (i.e. divulge and tryst) consider content generation for each corpora in collection as follows. Each content from corpora is drawn from one of themes, such that selected theme is taken from per-corpora distribution over themes. Thus, contexts proliferate from original content without any manual-labeling.

**Algorithm 1:** Yuva Algorithm
**Data:** Observed content, context
**Result:** Thematic representation of contexts
 begin Thematic scheme
   begin espouse
      Randomize theme(i.e.context) distribution
  For each content
    begin divulge
      Select randomized (i.e inspired by espouse) distribution
    begin tryst
        Select random content accordingly

## ASSESSMENT ON PROTOTYPE DEVELOPMENT

### Subjective Assessment

Survey is conducted among 77 participants about their perceptions on QoE in different emotions (i.e. happiness, sadness, boredom, relaxation), locations (i.e. workplace, subway, home), weather (i.e. cloudy, snowy, fair, sunny) and time (i.e. morning, lunch,evening, night), etc. Questionnaires are originally set to contemplate their perceptions in different contexts In the meantime, their fondness with Smart-devices are unveiled with some preliminary quiz. Media (e.g. Radio/TV) are noteworthy origins of multimedia traffic. Hence, they are asked about their opinions on media contents . Similarly, participants are adjured in sharing their social-network-attachments .

Assessments: Opinions are requested in a way, such that participants never turn confused. Hence, questionnaires are kept as simple as possible, which does not seem abstract to participants. In this context, several brainstorming sessions are undergone among co-authors (i.e. project members) to sustain a rhythm in the survey lifespan. The overall objective is to engage complete concentrations of participant, who promises to accomplish it with full motivation. Participants are encouraged to provide their perceptions within a short survey time. Because, human behavior changes in a short duration because of arrived phone-call, instant messaging or even request for an urgent work. Therefore, their feed-backs are gathered in leisure periods, while they are having idle time or just finished up with midterm exams, etc. Ample options are included in survey for the convenient remembrance of recent service-usage experiences. Human are fond of memorizing important or special moments from their past. It is scientifically proven that, participants have limitations in even remembrance of consecutive recent events. In this context, participants are asked about their perceptions separately, while they are traveling to/from workplace and at workplace. Participants are served with absolute ratings to quantify perception in conformity with literary descriptions. Similar experiences are interpreted by individuals in different ways. Moreover, psychological conditions have significant impact on personalized ratings. Therefore, user-ratings are divergent, even in the cases, when their perceptions are similar. In this context, survey questions are consigned for inferring emotional status and thereafter conjugal perceptions for service-usage. Perceptions are collected in versatile fashions. Because, participants are often at dis-comfortable zone in paying concentration on survey questionnaires. Such hasty opinions lack careful intention of participation. Therefore, trustworthiness of data is impaired due to such perfunctory obstacles. Such motivational reluctance is overcome with versatile and separate different contexts. Divergent data is collected from many participants to construct a meaningful representation of perceptions. Because, opinions from a small group of people give us abstract view about overall scenario. Collecting data from small group, with multiple opinions from a single person, is an alternative choice. However, repetition of participation could have negative impact on the quality of

collected data. Pr-assuming/keeping answers in mind from earlier leads to biased opinion or

**Objective Assessments**
Smart-device traffic is collected to monitor usage- information of lab members on different parts of a day. Usage information comprises of device identity (e.g. MAC address), application-genre (e.g. browsing), service-name (e.g. chrome) and time. Information is generally accumulated at the initiation of new session (e.g. opening new web-page/YouTube video/social feed ). Open source APIs and customized smart- device apps are synchronized to store information immediately in database. As monitoring operation is conducted inside the lab, outside weather has indirect influence on users experience. Due to the same reason, location information is static in the observation. Users anticipated perceptions (i.e. survey- feedback) about experiences on different locations (e.g. to- workplace, at workplace and from-workplace) is an alternative option at this point. Moreover, it is difficult to capture emotion through Smart-device traffic. Therefore, survey and traffic in- formation are quantified together to correlate user-experiences with psychological status
1) Assessments: Objective assessment entrenches high level of control in laboratory environment. Its archetypal outcome is rendered through communication patterns of individuals or groups. Participants use-profile validates this outcome to construct suitable models and parameters. These models are worthwhile to standardize methodologies for comparative analysis in real-life experiments. Subjective assessment is contingent on bizarre instincts, about which participant is unconscious. Therefore, it is harnessed as supplementary and statistical data source for pre- dictive models. Objective exposition of subjective assessment is merely quantitative, rather than qualitative. Thereby, quan- titative analysis blends objective and subjective counterparts in convenient decision making. Consequently, subjective one is frequently validated against objective counterpart to adjust unusual changes in quality of perceptions

**Acceptability Assessments**
Smart phone application is developed to accept QoE ratings from participants. Graphical user interface (GUI) is designed in a convenient way to annihilate negative impression on participant. Hence, questionnaires appear before and after application usage. Application demands following properties from participants.

-Psychological status prior to application- usage
-Psychological impact of recommendation

Assessments: Acceptability measures participants over- all experience on developed applications in online fashion. Subjective and objective assessments are conducted in limited space (i.e. subjects from only four communities) and indoor environment (i.e laboratory set-up), respectively. As a result, participants perception and satisfaction are seldom reflected, especially in real usage contexts. Hence, acceptability accumulates real-time feedback of participants in practical application scenarios. Thus, network, device and content-centric features are experienced in different dimensions through online and collaborative feed-backs. Engagement is articulated with network-centric (e.g. stalling) and device-centric(e.g. fractional download) features. On the contrary, termination is originated from network vulnerability, content quality or even unconscious mind. Unusual termination is fluctuated by participants profile and tolerance level. Network management-information gap (i.e. limited access on client-logs) is mitigable with the help of offline survey or real-time feedback. Content quality stems from resolution and application-awareness. Moreover, the integration of individual or social cultural contexts adheres content popularity. Immersiveness attracts participant with the thrill of virtual presence in digitized world. It is correlated with emotional, cultural, educational or environmental varieties. As a result, substantial reaction is accumulated on content having psychological attachments. Assessment among limited people interprets culture and personality in localized scale. This necessitates integration of generality for the betterment of quality and enjoyment in global scale.

**Validation of Assessments**
Participants are requested to compare two stimuli (Like,Unlike) on app and rate based on his/her own choice. However,in validation, participants are asked to reevaluate perception in a convenient group-discussion environment. Furthermore,multi scale-based rating (e.g. four-scale in acceptability) is shrinked to dichotomous choice, which eases decision making.
 Assessments: Cognitive gap between So So, Worst is different than that between Awesome and Good. Acceptability is rated according to arithmetic average of ratings, which is not applicable in ordinal scale measurements. Such precision limitations are overcome with dichotomous choice-based easier computation. Acceptability is impaired due to unconsciousness or reluctance of participants. Group evaluation overcome such problematic inputs from untrustworthy participation.

**CONCLUSION**
A dearth of adequate methodology prevail for elderly solicitude measurement in independent living. IoT-driven personalized services pave the way for improvement of such qualitative experiences. Eventually, context awareness is deemed both possibilities and pitfall for overall acceptability of such services. It avails plethora of participation information for the systematic realization of groundbreaking elderly services. However, the divergence in participation objectives introduces composite relationship between engagement and contexts. Even few pro-active and reactive factors have merely negative impact on engagement, rather than warmth acceptability. Ours is a first step in harnessing the blue ocean, while meeting major pitfalls. In this context, Topic-model based scheme resembles complex interactions among contexts, whilst scaling up over time. Consequently, prototype implementation is demonstrated for the accurate prediction of engagement by quantifying participants' opinions, usages or even feed-backs.